# Near-Field Thermophotovoltaic Energy Conversion with Thin-film Tandem Cells

Payam Sabbaghi, Qing Ni, and Liping Wang*

*School for Engineering of Matter, Transport, and Energy*

*Arizona State University, Tempe, AZ, 85287, USA*

* Corresponding author: <u>liping.wang@asu.edu</u>

**Abstract**

The performance of a near-field thermophotovoltaic system with a tandem-cell structure composed of thin-film *p*-doped GaSb and *n*-doped InAs sub-cells on a gold backside reflector is theoretically investigated. The temperatures of the Ga-doped ZnO emitter and the tandem cells are set as 1800 K and 300 K, while the thicknesses of GaSb and InAs sub-cells are considered as 1.5 μm and 0.5 μm, respectively. Fluctuational electrodynamics along with the multilayer dyadic Green's function is used to study near-field radiative heat transfer with the consideration of photon chemical potential, whereas radiative recombination and nonradiative Auger recombination are taken into account for evaluating the electrical performance of the tandem cells. At a vacuum gap of 50 nm, it is found that the tandem cells with independent charge collections achieve electrical power output 468.8 kW/m$^2$ at a conversion efficiency of 41%, generating relatively 87% (or 21%) more power with about absolute 5% (or 10%) higher efficiency than the single GaSb (or InAs) cell of the same 2-μm thickness. The physical mechanism of near-field spectral heat transfer is elucidated with energy transmission coefficient, while the current-voltage characteristics of sub-cells are discussed in detail. This work will pave the way to enhance near-field thermophotovoltaic energy conversion performance with tandem or multi-junction cells.

By bringing the emitter and the semiconductor cell into close proximity with distance much smaller than the thermal characteristic wavelength, near-field thermophotovoltaic (NF-TPV) systems could enhance the conversion performance from heat to electricity with the contribution of evanescent waves across the vacuum gap.[1-7] Recent studies have been focusing on the spectral control of near-field radiative spectrum in order to concentrate useful photons above cell bandgap for efficiency improvement by employing thin film,[8,9] multilayers,[10-12] gratings,[13-16] nanowire[17-20] metamaterial structures. Inoue et al.[21] proposed a one-chip NF-TPV device consisting of an intermediate Si substrate to suppress sub-bandgap radiation by free carriers and surface modes, yielding a 10-fold enhancement of the cell photocurrent compared to a far-field device. Stacking of multiple plasmonic thin film layers was also shown to simultaneously improve the efficiency and power output of NF-TPV systems.[22]

Another possible approach to improve the efficiency of a NF-TPV system is to harvest useful photons in a broader spectral range by using tandem or multi-junction cells, in which case the sub-cell with higher energy bandgap is placed on top to allow low-energy photons to pass and be absorbed by bottom sub-cell with lower bandgap. Note that tandem or multi-junction cells have been studied extensively for solar energy conversion in the far field.[23-29] Sahoo et al.[30] theoretically predicted a solar power conversion efficiency of 54% with a GaAs/GaSb dual-junction solar cell. Chen et al.[31] experimentally achieved record-high power conversion efficiency of 25.4% with a tandem solar cell made of silicon and halide perovskites. It has been also analyzed that stacking of individual cells into a tandem structure could increase the total electrical power output and conversion efficiency in solar TPV applications.[32-35] Wilt et al.[36] theoretically studied a monolithic interconnected modules by series-interconnected cells to reduce resistive losses and achieved higher output power density and system efficiency. However, the effect of tandem-cell structures has not been studied yet on near-field TPV systems.

In this study, we theoretically investigate the radiative heat transfer and power generation of a near-field thermophotovoltaic system by absorbing photons in a wide frequency range with the thin-film tandem cell structure, and demonstrate that the power output and conversion efficiency with tandem cell can greatly exceed those with a single cell for NF-TPV systems. As illustrated in Fig. 1(a), a thin layer of *p*-doped GaSb with a bandgap of $\omega_{g3} = 1.1 \times 10^{15}$ rad/s (i.e., 0.726 eV) is considered as the upper cell, while a thin-film *n*-doped InAs with a smaller bandgap of $\omega_{g4} = 0.54 \times 10^{15}$ rad/s (i.e., 0.354 eV) serves as the lower cell. It is expected for GaSb to absorb high-energy photons and for InAs to harvest low-energy ones, in order to convert thermal radiation in a wider frequency to electricity. The tandem cell is also supported by a gold backside reflector for photon recycling.[37] Both radiative recombination as a function of photon chemical potential and nonradiative Auger recombination are considered for the optoelectronic analysis. A plasmonic bulk emitter made of Ga-doped ZnO (GZO) is placed in close proximity to the tandem cells with deep subwavelength vacuum gap distance (i.e., $d = 50$ nm as a nominal value, which is possible to achieve experimentally[38]), such that the coupling of evanescent waves could contribute to super-Planckian radiative heat transfer and possibly enhance the energy conversion such as conversion efficiency and power output density.

The temperatures of the GZO emitter and the tandem cells are respectively fixed at $T_e = 1800$ K and $T_r = 300$ K, while the thicknesses of thin-film GaSb and InAs sub-cells are respectively set as $t_3 = 1.5$ μm, and $t_4 = 0.5$ μm. GZO is selected as the emitter material due to its high-temperature stability and strong plasmonic behaviors in the near-infrared matching the bandgap of cells. Dielectric function of GZO can be described by the Drude-Lorentz model as $\varepsilon_1(\omega) = \varepsilon_\infty - $ $\frac{\omega_{p1}^2}{\omega(\omega+i\Gamma_{p1})} + \frac{f_1\omega_1^2}{\omega_1^2-\omega^2-i\omega\Gamma_1}$ in which $\varepsilon_\infty = 2.475$, $\omega_{p1} = 1.927$ eV, $\Gamma_{p1} = 0.117$ eV, $f_1 = 0.866$ eV, $\omega_1 = 4.850$ eV and $\Gamma_1 = 0.029$ eV.[16,39] The upper GaSb cell is considered as *p*-doped with a doping level of $1.3\times10^{17}$ cm$^{-3}$,[40] and its dielectric function is obtained as the superposition from Palik's tabulated data with bandgap absorption and an analytical Drude term as $\varepsilon_3(\omega) = -15.1\frac{\omega_{p3}^2}{\omega^2+i\Gamma_{p3}\omega}$ to capture the contribution of free carriers from the doping ($\omega_{p3} = 0.017$ eV and $\Gamma_{p3} = 0.014$ eV),[40] which modifies the optical response mainly at the frequencies below $3\times10^{14}$ rad/s. The lower InAs cell is lightly *n*-doped with a doping level of $2\times10^{16}$ cm$^{-3}$,[41] and the dispersive dielectric function of InAs is taken from Palik's tabulated data of intrinsic one[42] with doping effect neglected. The optical properties of the gold back reflector is described by a Drude model with parameters from Ref. 43. The real and imaginary parts of dielectric functions of GZO, *p*-doped GaSb, *n*-doped InAs, and gold reflector are shown in Fig. S1 (see Supplementary Materials).

The near-field spectral radiative heat flux emitted by the GZO emitter that reaches each sub-cell and gold reflector across vacuum gap *d* can be calculated by fluctuational electrodynamics with the dyadic Green's function for isotropic homogeneous multilayers in the expression as[44]

$$q_{\omega,e\to r_j} = \frac{2k_0^2\Theta_e(\omega,T_e)}{\pi} \int_d^\infty Re[i\sum_{i=x,y,z}\varepsilon_1''(\omega)(G_{xi}^E G_{yi}^{H*} - G_{yi}^E G_{xi}^{H*})]\, dz \qquad (1)$$

where $j = 3$, 4 or 5 indicates the GaSb, InAs or gold layer, $k_0$ is the wavevector in vacuum, and $\varepsilon_1''$ is the imaginary part of the dielectric function of the GZO emitter. $\Theta_e(\omega,T_e) = \hbar\omega/\left[exp(\frac{\hbar\omega}{k_B T_e}) - 1\right]$ represents the mean energy of a Planck oscillator for the emitting layer at temperature $T_e$, where $\hbar$ is the reduced Planck constant and $k_B$ is the Boltzmann constant. The spectral radiative heat flux from each sub-cell to the emitter can be obtained in a similar form as

$$q_{\omega,r_j\to e}(V_j) = \frac{2k_0^2\Theta_{r,j}(\omega,T_r,V_j)}{\pi} \int_{z_j}^{z_{j+1}} Re[i\sum_{i=x,y,z}\varepsilon_j''(\omega)(G_{xi}^E G_{yi}^{H*} - G_{yi}^E G_{xi}^{H*})]\, dz \qquad (2)$$

Note that the mean energy of a Planck oscillator for each sub-cell considers the effect of photon chemical potential due to biasing with a modified expression as[45-48]

$$\Theta_{r,j}(\omega, T_r, V_j) = \hbar\omega / \left[\exp(\frac{\hbar\omega - eV_j}{k_B T_r}) - 1\right] \qquad (3)$$

where $e$ is the elementary charge, and $V_j$ is the applied voltage on each sub-cell. Therefore, the net spectral radiative heat flux from the GZO emitter to each GaSb or InAs sub-cell should be $q_{\omega,r_j} = q_{\omega,e \to r_j} - q_{\omega,r_j \to e}(V_j)$, and that from the emitter to the entire cell structure is $q_{\omega,r} = \sum_{j=3,4,5} q_{\omega,r_j}$. Note that with high emitter temperature of 1800 K considered here, $q_{\omega,r_j \to e}(V_j)$ is negligibly small compared to $q_{\omega,e \to r_j}$ such that the net spectral radiative heat flux $q_{\omega,r_j}$ and $q_{\omega,r}$ can be reasonably considered independent of bias voltages of sub-cells (see Fig. S2).

Figure 1(b) shows the calculated spectral radiative heat flux respectively absorbed by 1.5-μm-thick GaSb and 0.5-μm-thick InAs cell in the tandem structure at a vacuum gap of 50 nm with the GZO emitter at 1800 K. Clearly, the upper GaSb cell absorbs photons greatly from its bandgap of $\omega_{g3} = 1.1 \times 10^{15}$ rad/s to $2.5 \times 10^{15}$ rad/s in a narrowband peak with maximum spectral heat flux value about 1 nJ·m⁻²·rad⁻¹ at the angular frequency of $1.3 \times 10^{15}$ rad/s. On the other hand, the bottom InAs cell mainly harvests the photons with lower energy between the two bandgaps (i.e., $\omega_{g3}$ to $\omega_{g4}$) broadly with high flux between 1 and 1.2 nJ·m⁻²·rad⁻¹ due to multiple spectral peaks. At frequencies lower than $\omega_{g4} = 0.54 \times 10^{14}$ rad/s where photons are not useful to generate power, the thin-film InAs cell barely absorbs but the GaSb cell exhibits small spectral radiative flux around 0.2 nJ·m⁻²·rad⁻¹ due to free-carrier absorption from its slightly higher doping level. In comparison, the spectral near-field radiative heat flux of a single cell of either *p*-doped GaSb or *n*-doped InAs with the same 2-μm thickness of the tandem structure and backside gold reflector was also presented. While the single cell shows slightly higher peak spectral radiative heat flux (i.e., 1.2

nJ·m$^{-2}$·rad$^{-1}$ for single 2-μm-thick GaSb cell and 1.4 nJ·m$^{-2}$·rad$^{-1}$ for single 2-μm-thick InAs cell) mainly due to the greater cell thickness when compared to that in the tandem structure. By employing a tandem structure, the absorbed radiative heat flux or useful thermal photons for power generation are available in a broader spectral range than the single-cell counterpart. Interestingly, the single GaSb cell also shows a narrowband absorption, whereas the single InAs cell displays a broadband spectral heat flux, similar to those observed in the tandem-cell structure.

In order to understand the physical mechanisms of near-field radiative heat transfer between the emitter and tandem-cell structure across the 50-nm vacuum gap, the energy transmission coefficient $\xi_j(\omega,\beta)$ for random thermal emission including both *s* and *p* polarizations from the GZO emitter to the each sub-cell *j* is obtained from

$$q_{\omega,e \to r_j} = \frac{\Theta_e(\omega,T_e)}{4\pi^2} \int_0^\infty \beta \xi_j(\omega,\beta) \, d\beta \tag{4}$$

As shown in Figs. 2(a) and 2(b), the energy transmission coefficient $\xi_j$ for the 1.5-μm-thick GaSb sub-cell and 0.5-μm-thick InAs sub-cell, is respectively plotted as a function of frequency $\omega$ and in-plane wavevector $\beta$ normalized by $\omega_0/c_0$ with $\omega_0 = 10^{14}$ rad/s and $c_0$ is the speed of light in vacuum. In addition, the light lines inside the vacuum and sub-cells are also plotted respectively from $\beta = \omega/c_0$ and $\beta = n_j\omega/c_0$, where $n_j = Re(\sqrt{\varepsilon_j})$ is the refractive index of either GaSb (*j* =3) or InAs (*j* =4) sub-cell. Clearly, enhanced radiative transfer indicated by the bright contour with high energy transmission coefficient close to 2 can be observed from both sub-cells only in the areas below the vacuum light line as propagating modes (i.e., $0 < \beta c_0/\omega < 1$) and between the vacuum and sub-cell light lines as frustrated modes (i.e., $1 < \beta c_0/\omega < n_j$). Moreover, the Fabry-Perot-like enhancement exhibits multiple fringes suggesting strong interference effect that occurs inside the thin-film sub-cell layer, which was also observed in single thin-film cell for NF-TPV but named "thermal-well" effect.[8] As seen in Fig. 2(b), the interference effect is more obvious for *n*-type InAs due to its smaller thickness, which causes the multiple peaks in the spectral heat flux in a broad band from to 0.55 to $1.2\times10^{15}$ rad/s. On the other hand in Fig. 2(a), the heat transfer enhancement occurs at higher frequencies for the 1.5-µm GaSb sub-cell due to its larger bandgap and peaks at large $\beta$ values around $1.3\times10^{15}$ rad/s, which leads to a narrowband peak in the spectral radiative flux.

In order to evaluate the electrical performance of the tandem cell, the current density inside each sub-cell can be calculated as[41]

$$I_j = e[F_{1j} - F_{j1}(V_j) - R_j(V_j)] \tag{5}$$

Note that $F_{1j}$ denotes the electron-hole pair generation rate (per unit volume) by absorbing the photons with energy greater than the bandgap of sub-cell $j$ emitted from medium 1 as

$$F_{1j} = \frac{2}{\pi}\int_{\omega_{g,j}}^{\infty}\frac{k_0^2\Theta_e(\omega,T_e)}{\hbar\omega}\{\int_d^{\infty}Re[\,i\sum_{i\,=\,x,y,z}\varepsilon_1''(\omega)(G_{xi}^E G_{yi}^{H*} - G_{yi}^E G_{xi}^{H*})]\;dz\}d\omega \tag{6}$$

and $F_{j1}(V_j)$ represents bias-dependent radiative recombination rate (per unit volume) by emitting the photons from sub-cell $j$ to the GZO emitter (i.e., medium 1) as

$$F_{j1}(V_j) = \frac{2}{\pi}\int_{\omega_{g,j}}^{\infty}\frac{k_0^2\Theta_{r,j}(\omega,T_r,V_j)}{\hbar\omega}\{\int Re[\,i\sum_{i\,=\,x,y,z}\varepsilon_j''(\omega)(G_{xi}^E G_{yi}^{H*} - G_{yi}^E G_{xi}^{H*})]\;dz\}d\omega \tag{7}$$

and $R_j(V_j)$ is the nonradiative Auger recombination rate inside the sub-cell $j$ calculated by[41]

$$R(V) = (C_n n + C_p p)(np - n_0^2)t \tag{8}$$

where $n$ and $p$ are respectively the electron and hole concentrations depending on cell voltage and the doping level, $C_p$ and $C_n$ represent the Auger recombination coefficients, and $n_0$ is the intrinsic carrier concentration, whose values are taken from Refs. [40,41] for GaSb and InAs sub-cells.

Figure 3 shows the calculated current-voltage curves of the 1.5-µm-thick GaSb and 0.5-µm-thick InAs sub-cells in the tandem structure along with that from their single-cell counterpart with 2-µm thickness, as a result of near-field radiative heat transfer from the GZO emitter at 1800

K across the 50-nm vacuum gap. The short-circuit current density, which is dominated by the photon-excited free-carrier generation (i.e., $F_{1j}$), is $I_{sc,3} = 40.2$ A/cm$^2$ for the *p*-doped GaSb upper sub-cell from the tandem structure, only 14% smaller than 46.6 A/cm$^2$ from the 25% thicker single-cell GaSb structure. While the bottom InAs sub-cell from the tandem structure is only 25% thick of its single-cell counterpart, it reaches a short-circuit current $I_{sc,4} = 117.5$ A/cm$^2$, producing more than 60% of four-time-thick single cell. On the other hand, the open-circuit voltage where the current is zero is found to be $V_{OC,3} = 0.64$ V for the GaSb cell and $V_{OC,4} = 0.25$ V for the InAs cell regardless of single- or tandem-cell structure. Note that open-circuit voltage is mainly affected by the photoluminescence quantum yield (PLQY), which is the rate of radiative recombination divided by the total (radiative and nonradiative) recombination. A higher nonradiative recombination rate or a smaller PLQY value reduces open-circuit voltage.[49] With additional calculations of each term in Eq. 4 as a function of sub-cell voltage (see Figs. S3), the PLQY is 93% and 11% respectively for the GaSb and InAs sub-cells in the tandem structure, indicating that radiative recombination dominates in the GaSb sub-cell and the nonradiative Auger recombination dominates in the InAs sub-cell.

The total power generation from the tandem cell depends on the particular sub-cell electric architecture (i.e., in-series, parallel, and independent). Here an independent charge collection is considered in this work for simplicity to demonstrate the electric performance of the NF-TPV with tandem cell, in which case the total electric power density is the summation of maximal values from each sub-cell as

$$P_{\text{total}} = \text{Max}(I_3 V_3) + \text{Max}(I_4 V_4) \tag{9}$$

The heat-to-power conversion efficiency of the NF-TPV system with tandem-cell structure is

$$\eta = \frac{P_{\text{total}}}{q_{\text{total}}} \tag{10}$$

where $q_{total} = \int q_{\omega,r} d\omega$ is the total radiative heat flux absorbed by both sub-cells and the gold backside reflector.

From the calculated current-voltage curves in Fig. 3, the maximum powers respectively produced by the 1.5-μm-thick GaSb and 0.5-μm-thick InAs sub-cells are $P_{3,max}$ = 219 kW/m² (at optimal $V_{3,opt}$ = 0.58 V and $I_{3,opt}$ = 37.8 A/cm²) and $P_{4,max}$ = 250 kW/m² (at $V_{4,opt}$ = 0.22 V and $I_{3,opt}$ = 113.4 A/cm²), resulting in a total power density $P_{total}$ = 469 kW/m² from the tandem-cell structure. In comparison, the 2-μm-thick single-cell counterpart alone produces $P'_{3,max}$ = 251 kW/m² with GaSb or $P'_{4,max}$ = 388 kW/m² with InAs, at respectively the same optimal $V_{oc}$ but different $I_{sc}$ values. Apparently, the tandem cell produces relatively 87% or 21% more power than the single GaSb or InAs cell of the same thickness. On the other hand, the total near-field radiative heat flux $q_{total}$ is 1145 kW/m² on the tandem cell, 690 kW/m² on the single GaSb cell, and 1248 kW/m² on the single InAs cell, all of which are supported by gold reflectors. As a result, the conversion efficiency of a NF-TPV system with the GZO emitter at 1800 K across 50-nm vacuum gap is found to be 40.9% from the tandem cell, demonstrating an (absolute) improvement of 4.6% or 9.8% compared to GaSb or InAs single cell of the same total thickness.

Finally, the total electrical power output and conversion efficiency of the NF-TPV systems with either tandem- or single-cell structures are calculated at a few selected vacuum gap distances from 500 nm down to 20 nm. As shown in Fig. 4(a), the electric power monotonically increases as the vacuum gap distance decreases from the contribution of evanescent waves in the near-field regime for all cases. At vacuum gap of 50 nm or larger, the tandem cell produces the most power while the InAs or GaSb single cell generates up to 20% or 50% less power than the tandem cell. Interestingly at vacuum gap $d$ = 20 nm, the InAs single cell actually produces 868 kW/m² power, 9% more than the tandem one (796 kW/m²) and 135% more than the GaSb single cell (369 kW/m²).

As a matter of fact, as shown in Fig. 4(b), the tandem-cell structure exhibits higher conversion efficiency at all vacuum gaps, in particular, achieving 42.5% at $d = 20$ nm, which is (absolute) 9.8% or 5.4% higher than either InAs or GaSb single-cell counterpart. Note that the analysis here is based on the tandem cell with 1.5-µm-thick GaSb upper cell and 0.5-µm-thick InAs bottom cell in comparison with single cell of the same 2-µm total thickness. The conversion performance improvement with the tandem-cell structure could be possibly greater by optimizing the sub-cell thickness ratio.

In summary, we have theoretically studied the performance of a NF-TPV system with thin-film tandem-cell structure which could convert useful photons in a wider spectral range into electricity. At a vacuum gap distance of 50 nm with GZO emitter at 1800 K, the tandem cell consisting of individually connected 1.5-µm GaSb and 0.5-µm InAs sub-cells produces electric power of 469 kW/m$^2$ at 40.9% conversion efficiency, demonstrating the improved performance over the single cell of GaSb (251 kW/m$^2$ and 36.3%) or InAs (388 kW/m$^2$ and 31.1%) with the same 2-µm thickness. It is found that the GaSb sub-cell has higher open-circuit voltage than the InAs one due to radiative loss dominating the recombination with high photoluminescence quantum yield. The performance of NF-TPV system with tandem-cell structure could be possibly further improved by optimizing the sub-cell thickness-ratio, material combinations for the emitter, sub-cells, and coatings, sub-cell electrical connections, or more junctions with higher cost and complexity.[50] This work opens a pathway to develop more efficient near-field TPV systems by employing the tandem-cell structures.

**Acknowledgements**

This work was supported by National Science Foundation (Grant No. CBET-1454698) and Air Force Office of Scientific Research (Grant No. FA9550-17-1-0080). P.S. would also like to thank ASU Graduate College for providing the Completion Fellowship.

**Data Availability**

The data that support the findings of this study is available from the corresponding author upon reasonable request.

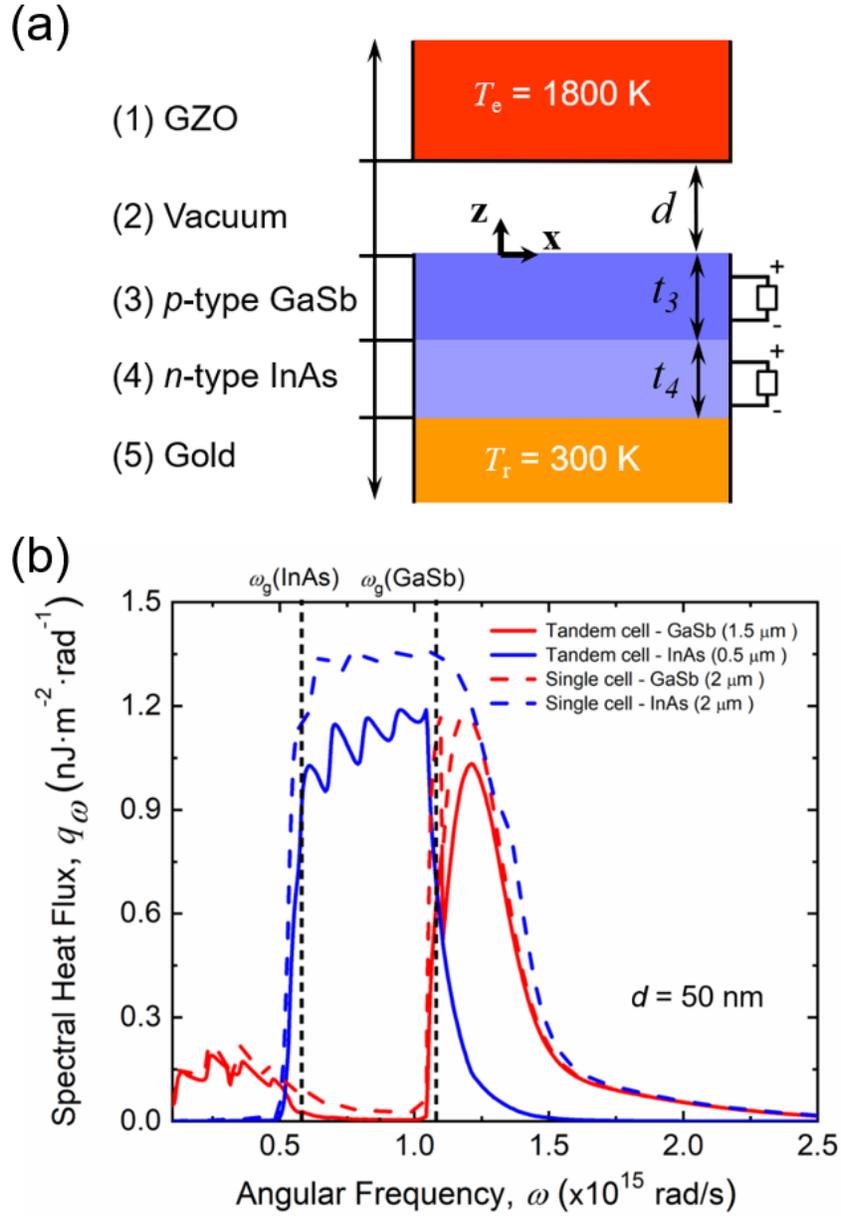

Figure 1. (a) Schematic of a near-field TPV system with thin-film tandem cells consisting of *p*-doped GaSb and *n*-doped InAs as the upper and lower sub-cells with a vacuum gap distance *d*. The temperatures of the GZO emitter and the tandem cells are fixed at 1800 K and 300 K, respectively. (b) Calculated near-field spectral radiative heat flux absorbed by each GaSb or InAs sub-cell in the dual-junction tandem structure in comparison with that absorbed by a single-junction cell with the same thickness. The total thickness of the tandem or single cell structure is 2 μm while the thickness of first and second cells are respectively set as $t_3$ = 1.5 μm and $t_4$ = 0.5 μm, and the vacuum gap spacing is *d* = 50 nm.

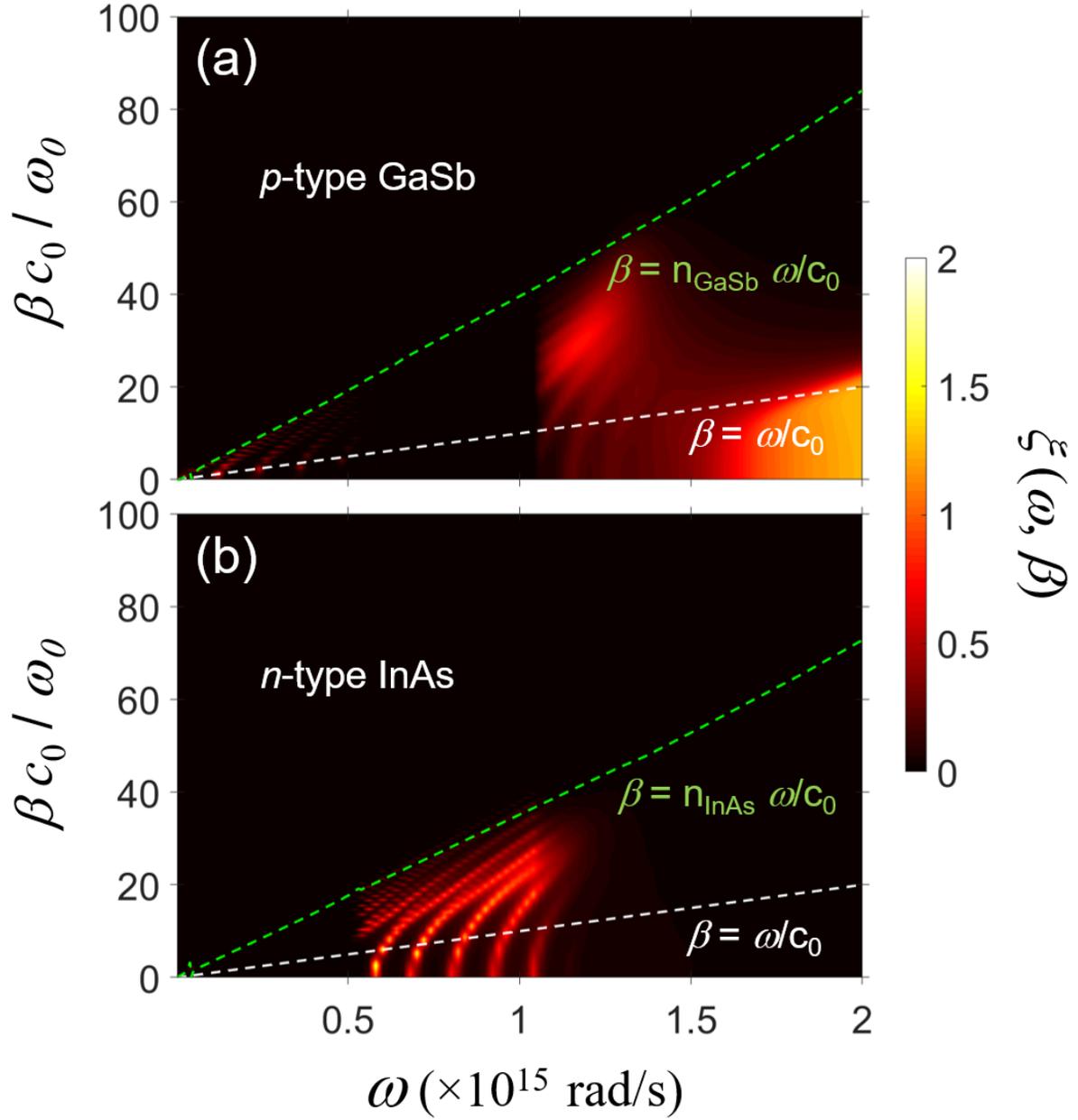

Figure 2. Contour plot of energy transmission coefficient ($\xi$) for each sub-cell in the tandem structure for both *s* and *p* polarizations at $d = 50$ nm vacuum gap distance: (a) *p*-type GaSb sub-cell and (b) *n*-type InAS sub-cell. Note that the parallel wavevector component is normalized by $\beta_0 = \omega_0/c_0$ with $\omega_0 = 10^{14}$ rad/s and $c_0$ is the speed of light in vacuum.

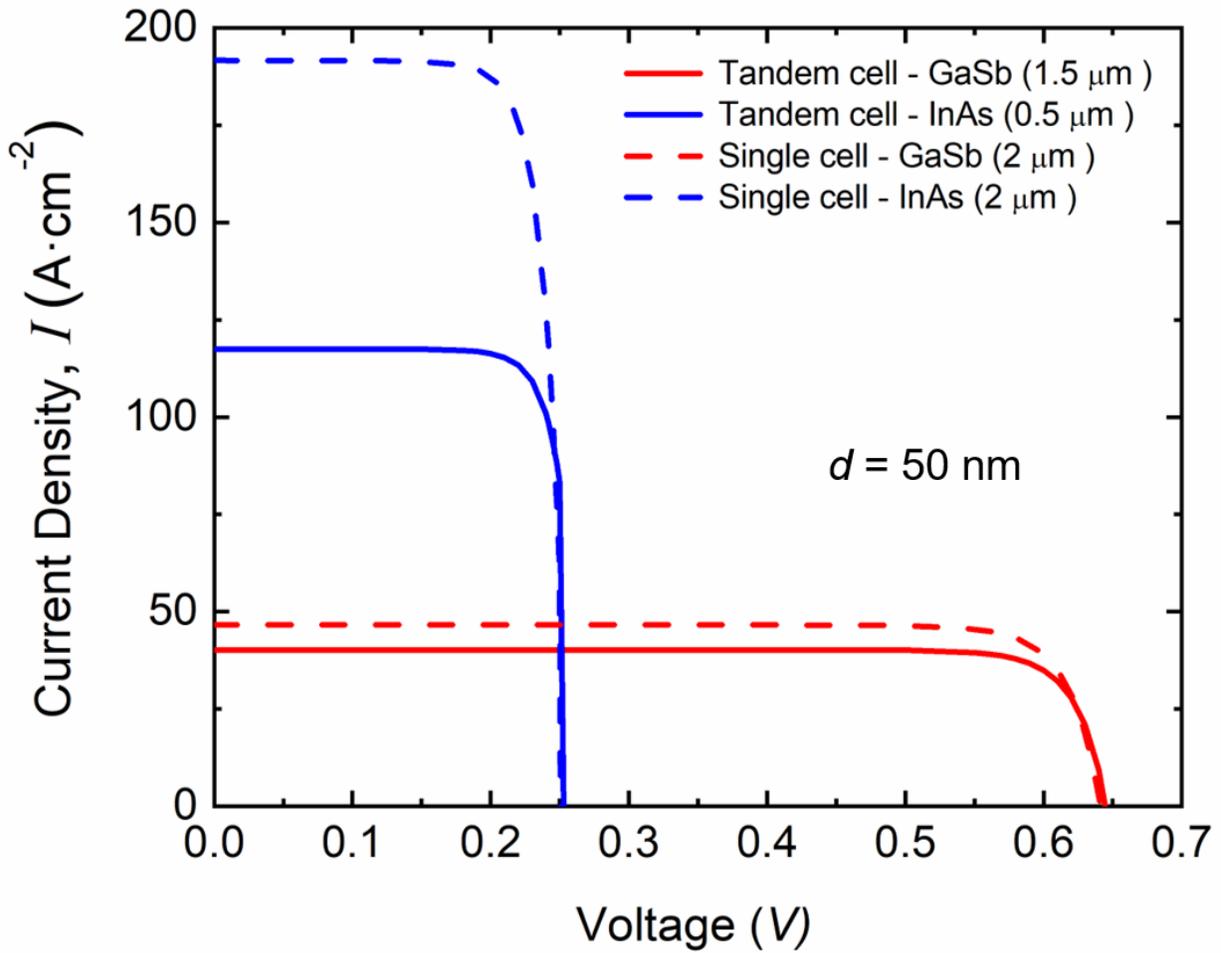

Figure 3. The current-voltage characteristics of each GaSb (1.5 μm thick) or InAs (0.5 μm thick) sub-cell in the dual-junction tandem structure in comparison with that of a single-junction cell with the same 2-μm thickness at the vacuum gap spacing $d = 50$ nm.

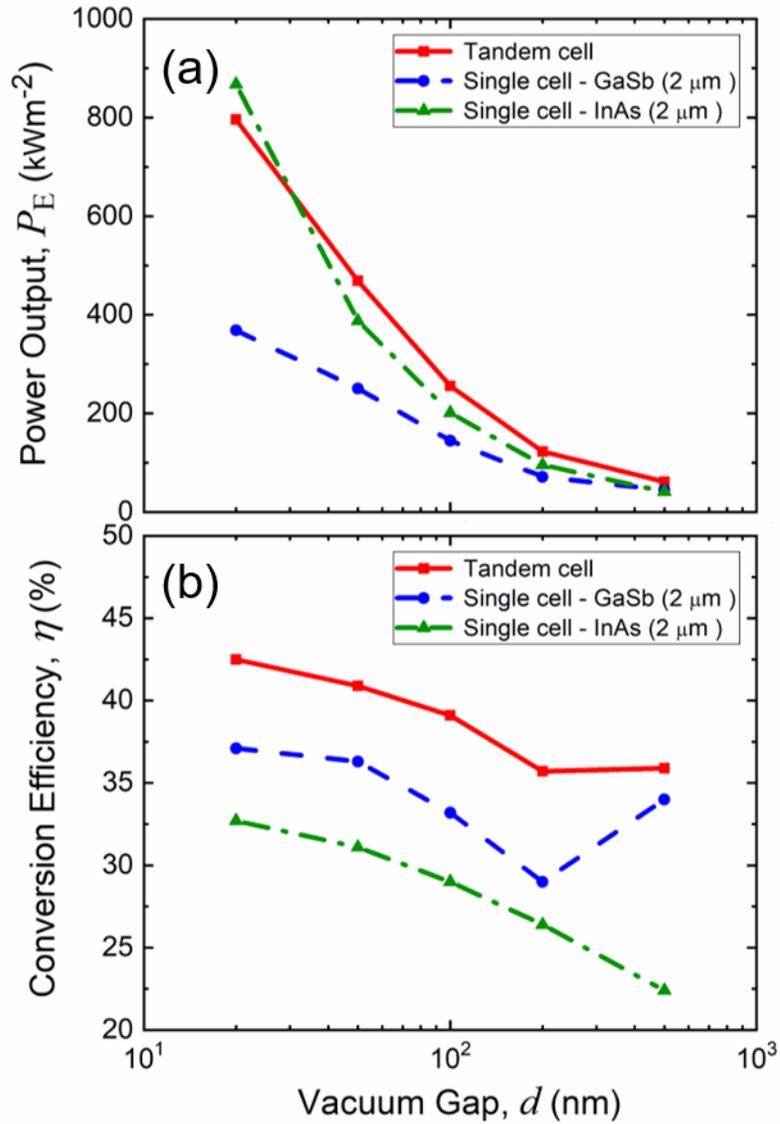

Figure 4. The effect of vacuum gap distance on (a) electrical power generation, and (b) conversion efficiency for a near-field TPV system with the tandem structure consisting of 1.5-μm-thick GaSb and 0.5-μm-thick InAs sub-cells in comparison to those with single-junction cell of the same 2-μm thickness.